\documentclass[12pt]{article}

\usepackage[margin=1in]{geometry}

\usepackage{graphicx}
\usepackage{natbib}
\usepackage{amsmath,amssymb}

\usepackage{authblk}

\usepackage{hyperref}

\usepackage{chngcntr} 
\newcommand{\beginsupplement}{%
  \setcounter{section}{0}%
  \setcounter{figure}{0}%
  \setcounter{table}{0}%
  \setcounter{equation}{0}%
  \renewcommand{\thesection}{S\arabic{section}}%
  \renewcommand{\thefigure}{S\arabic{figure}}%
  \renewcommand{\thetable}{S\arabic{table}}%
  \renewcommand{\theequation}{S\arabic{equation}}%
}

\title{Characterization of low-nitrogen quantum diamond for pulsed magnetometry applications}

\author[1, 2]{Jiashen Tang}
\author[5]{Connor A. Roncaioli}
\author[6]{Andrew M. Edmonds}
\author[4]{Atli Davidsson}
\author[1]{Connor A. Hart}
\author[6]{Matthew L. Markham}
\author[1, 2, 3]{Ronald L. Walsworth\thanks{Correspondence: Ronald L. Walsworth (walsworth@umd.edu).}}

\affil[1]{Quantum Technology Center, University of Maryland, College Park, Maryland, USA}
\affil[2]{Department of Physics, University of Maryland, College Park, Maryland, USA}
\affil[3]{Department of Electrical Engineering and Computer Science, University of Maryland, College Park, Maryland, USA}
\affil[4]{Department of Chemistry and Biochemistry, University of Maryland, College park, Maryland, USA}
\affil[5]{DEVCOM Army Research Laboratory, Adelphi, Maryland, USA}
\affil[6]{Element Six Global Innovation Centre, Fermi Avenue, Harwell Oxford, Didcot, Oxfordshire, United Kingdom}

\date{}

\begin{document}
\maketitle

\begin{abstract}
Ensembles of nitrogen-vacancy (NV) centers in diamond are versatile quantum sensors with broad applications in the physical and life sciences. The concentration of neutral substitutional nitrogen ([N$_\text{s}^0$]) strongly influences coherence times, sensitivity, and optimal sensing strategies. Diamonds with [N$_\text{s}^0$] $\sim$\,1-10\,ppm are a focus of recent material engineering efforts, with higher concentrations being favorable for continuous-wave optically detected magnetic resonance (CW-ODMR) and lower concentrations expected to benefit pulsed magnetometry techniques through extended NV electronic spin coherence times and improved sensing duty cycles. In this work, we synthesize and characterize low-[N$_\text{s}^0$] ($\sim$\,0.8\,ppm), NV-enriched diamond material, engineered through low-strain chemical vapor deposition (CVD) growth on high-quality substrates, $^{12}$C isotopic purification, and controlled electron irradiation and annealing. Our results demonstrate good strain homogeneity in diamonds grown on CVD substrates and spin-bath-limited NV dephasing times. By measuring NV spin and charge properties across a wide range of optical NV excitation intensity, we provide direct comparisons of photon-shot-noise-limited magnetic sensitivity between the current low-[$\text{N}_\text{s}^0$] and previously studied higher-[$\text{N}_\text{s}^0$] ($\sim$\,14\,ppm) NV-diamond sensors. We show that low-[N$_\text{s}^0$] diamond can outperform higher-[N$_\text{s}^0$] diamond at moderate and low optical NV excitation intensity. Our results provide practical benchmarks and guidance for selecting NV-diamond sensors tailored to specific experimental constraints and sensing requirements.
\end{abstract}

\noindent 

\section{Introduction}

Nitrogen-vacancy (NV) centers in diamond are a leading quantum sensing platform due to favorable optical and electronic spin properties \citep{barry2020report}, with wide-ranging applications, including in condensed matter physics \citep{ku2020imaging, zhang2021ac, rovny2024nanoscale}, electronics systems \citep{Turner2020IC, garsi2024three}, geoscience \citep{glenn2017micrometer, fu2023pinpointing}, and life science \citep{barry2016optical,aslam2023quantum,schirhagl2014nitrogen}. Extensive research has focused on magnetic sensing and imaging using ensembles of NV centers. However, experimentally realized sensitivities remain several orders of magnitude below the fundamental spin-projection limit \citep{Hart2021-4Ramsey,barry2024sensitive}. A recent review \citep{barry2020report} highlights strategies for further improving sensitivity, including extending spin coherence times through advanced pulse sequences and targeted material engineering.

A key material engineering parameter affecting NV ensemble magnetic sensing performance is the concentration of neutral substitutional nitrogen ([N$_\text{s}^{0}$]), which strongly influences NV formation, charge stability, and spin coherence properties. Diamonds with higher [N$_\text{s}^{0}$] provide increased NV center density and improved NV$^-$ charge stability due to electron-donating nitrogen. However, higher [N$_\text{s}^0$] also creates a dense paramagnetic spin bath, significantly limiting NV spin coherence times. Conversely, lower [N$_\text{s}^{0}$] materials offer improved coherence at the expense of NV density and charge stability. Previous studies identified an optimal [N$_\text{s}^{0}$] range around 1–10\,ppm for balancing these competing factors \citep{bauch2018ultralong, edmonds2021characterisation}.

\begin{figure}[htbp]
    \centering
    \includegraphics[width=\linewidth]{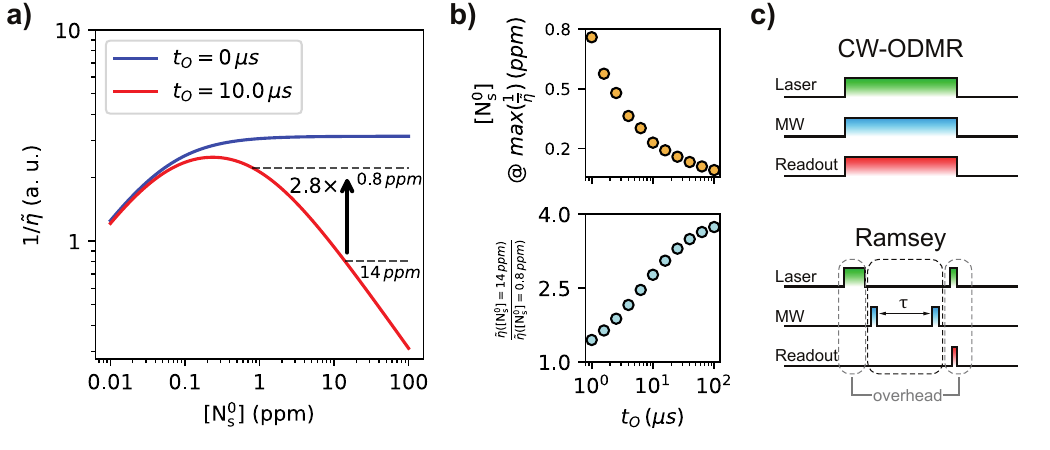}
    \caption{(a) NV ensemble DC magnetic field sensitivity estimated using the simplified metric 
    $\smash{1/\tilde{\eta} = \sqrt{[\mathrm{N}_\mathrm{s}^0]\times T_2^*}}$, assuming spin-bath-limited $T_2^*$ and 99.995\% isotopic enrichment of $^{12}\mathrm{C}$. 
    Corrections accounting for experimental overhead time ($t_O$) are included by multiplying with $\smash{1/\sqrt{(T_2^*+t_O)/T_2^*}}$ (cf. Eq.~\ref{eq:1}). 
    With zero overhead (blue line), sensitivity plateaus for $\smash{[\mathrm{N}_\mathrm{s}^0]\gtrsim 1\,\text{ppm}}$. 
    However, incorporating an overhead of $10\mu\, \text{s}$, typical in pulsed Ramsey protocols, shifts the optimal $\smash{[\mathrm{N}_\mathrm{s}^0]}$ towards lower values (red line). 
    A preliminary comparison between the current low-$\smash{[\mathrm{N}_\mathrm{s}^0]}$ samples ($\sim 0.8\,\text{ppm}$) and a previously characterized high-$\smash{[\mathrm{N}_\mathrm{s}^0]}$ sample ($\sim 14\,\text{ppm}\smash{}$ suggests a nearly threefold sensitivity improvement.
    (b) Estimated optimal nitrogen concentration, and sensitivity improvement between $0.8\,\text{ppm}\smash{}$ and $14\,\text{ppm}\smash{}$ samples as functions of overhead time ($t_O$).
    (c)  Schematics comparing NV magnetometry protocols based on continuous-wave optically detected magnetic resonance (CW-ODMR) and pulsed Ramsey. CW-ODMR has nominally zero overhead but suffers from intrinsic competition between simultaneous microwave (MW) and optical NV excitation. Pulsed Ramsey avoids these competing effects but incurs non-negligible overhead from NV initialization and readout.}
    \label{fig:1}
\end{figure}

In prior work \citep{edmonds2021characterisation}, we synthesized chemical vapor deposition (CVD) diamonds with relatively high [N$_\text{s}^0$] (nominally 14\,ppm), demonstrating a favorable balance between NV density and Ramsey dephasing time $T_2^*$ for broadband DC magnetic sensing. Shorter $T_2^*$ also reduces the demand on bias magnetic field homogeneity. However, sensitivity assessments using simplified metrics like $\tilde{\eta}=(\text{[N}_\text{s}^0\text{]}\times T_2^*)^{-1/2}$ inadequately capture experimental overheads from NV spin initialization and readout. Such overheads particularly impact pulsed magnetometry protocols, which are advantageous for high-sensitivity measurements due to their separation of microwave and optical broadening effects inherent in continuous-wave optically detected magnetic resonance (CW-ODMR) methods \citep{barry2020report}. Accounting for these overheads shifts the sensitivity optimum toward lower [N$_\text{s}^0$] diamonds (Fig.~\ref{fig:1}), where longer coherence improves sensing duty cycles.

Realizing these sensing advantages with low-nitrogen diamond materials necessitates careful control of other material parameters. These include $^{12}$C isotopic purification to minimize $^{13}$C nuclear spin noise, optimization of growth conditions to reduce strain and other unwanted defect densities, and controlled irradiation and annealing processes to achieve high N$_\text{s}^0$-to-NV conversion while preserving the NV$^-$ charge fraction despite lower donor concentrations.

In this work, we report the synthesis and characterization of low-[N$_\text{s}^0$] ($\sim$\,0.8\,ppm) diamond material for pulsed NV ensemble magnetometry. Focusing on Ramsey-based DC magnetic field sensing, we show that combining low-strain CVD growth, $^{12}$C isotopic purification, and controlled electron irradiation and annealing yields NV-enriched diamond materials with spin-bath-limited dephasing times. We directly compare photon-shot-noise-limited sensitivity between an example low-[N$_\text{s}^0$] sample and a previously studied high-[N$_\text{s}^0$] diamond \cite{edmonds2021characterisation}, by characterizing NV spin and charge properties across a range of laser intensities used for optical NV excitation (i.e., NV spin initialization and readout). The low-nitrogen sample demonstrates improved sensitivity, particularly at moderate and low optical powers. This study provides practical benchmarks and guidance for selecting NV diamond sensors across diverse magnetometry applications, considering experimental constraints and sensing requirements, and also informs future diamond material engineering efforts.

\section{Material and Experimental Methods}
\subsection{Parameters Relevant to Sensitivity}
The photon shot-noise-limited magnetic sensitivity of an NV ensemble Ramsey-based DC magnetometry protocol is given by:
\begin{equation}
    \eta_{Ramsey}^{photon-shot} = \frac{1}{\Delta m_s\gamma_e}\frac{1}{\sqrt{N\tau}}\frac{1}{e^{-(\tau/T_2^*)^p}}\sqrt{1+\frac{1}{C^2n_{nvg}}}\sqrt{\frac{\tau+t_O}{\tau}},
\label{eq:1}
\end{equation}
where $\Delta m_s$ denotes the spin-state transition (1 for single-quantum and 2 for double-quantum transitions \citep{mamin2014multipulse}), $\gamma_e$ is the electron gyromagnetic ratio, $N$ is the total number of NV$^-$ contributing to the measurement, $\tau$ is the spin free-precession time during sensing, $T_2^*$ and $p$ characterize the NV spin dephasing time and Ramsey decay envelope shape, $C$ is the NV spin-state-dependent photoluminescence (PL) measurement contrast, $n_{avg}$ is the average PL photon number detected per NV$^-$ per measurement, and $t_O$ is the total experimental overhead time including initialization and readout.

$T_2^*$ for an NV ensemble can be expressed as the inverse sum of several common dephasing mechanisms \citep{bauch2018ultralong,barry2020report}:
\begin{equation}
    \frac{1}{T_2^*}=\frac{1}{T_{2,\text{[N}_\text{s}^0\text{]}}^*}+\frac{1}{T_{2,\text{[}^{13}\text{C]}}^*}+\frac{1}{T_{2,\text{NV-NV}}^*}+\frac{1}{T_{2,\text{ strain}}^*}+\frac{1}{T_{2, \text{bias}}^*}.
    \label{eq:2}
\end{equation}
The first two terms represent the spin-bath noise from substitutional nitrogen and residual $^{13}$C in the diamond lattice, with their contributions scaling directly with respective concentrations. The third term accounts for NV–NV dipolar interactions, which can be significant in high NV density samples created by irradiation and annealing. The last two terms account for spatial variation (inhomogeneity) in lattice strain and the applied bias magnetic field. Quantification of [N$_\text{s}^0$], [NV], $^{13}$C content, and strain is therefore essential for confirming whether a given diamond sample achieves spin-bath-limited dephasing.

The NV optical excitation intensity further affects the photon-shot-noise-limited sensitivity by modifying the NV$^-$ charge fraction, and hence the effective number of sensors $N$. Since NV$^-$ and NV$^0$ have overlapping emission spectra and different PL brightness, both the measurement contrast ($C$) and detected photon number per NV$^-$ ($n_{\text{avg}}$) vary with excitation laser intensity. Additionally, the experimental overhead time $t_O$, which includes NV spin initialization and readout, also depends on optical NV excitation intensity. Therefore, in addition to measuring $T_2^*$, it is important to characterize these parameters as functions of excitation intensity to enable a complete sensitivity assessment.

\subsection{Diamond Synthesis and Treatment}
\label{sec:2_2}
The diamond samples studied here are grown on a diamond substrate by Element Six using microwave-plasma-assisted chemical vapor deposition (CVD). Nitrogen-doped CVD diamond materials can exhibit an increased fraction of positively charged substitutional nitrogen ([N$_\text{s}^+$]), associated with undesirable brown coloration. This coloration increases absorption of excitation light and NV PL, and is correlated with elevated concentrations of paramagnetic defects related to vacancies \citep{khan2010,khan2009,hounsome2006,fujita2009} and hydrogen \citep{glover2003,glover2004}. To mitigate these effects and achieve a homogeneous distribution of nitrogen with [N$_\text{s}^0$]\,$\approx 1$\,ppm, key synthesis parameters -- including substrate temperature, CH$_4$/H$_2$ ratio, and N$_2$ concentration -- are optimized. Additionally, CH$_4$ sources isotopically enriched with $^{12}$C ($>$\,99.99\%) are used. Deposition conditions are controlled for the duration of the run to avoid the formation of non-epitaxial crystallites on the main growth surface; and twinning at the edges of the samples is also minimized to avoid a reduction in usable surface area for the final processed plates.

Crystal strain in CVD-grown diamond layers is strongly influenced by the dislocation density, and dislocations typically propagate along the growth direction \citep{martineau2004,gaukroger2008}. Key sources of dislocations include those nucleated at the substrate–growth interface due to substrate polishing damage, and dislocations threading from the substrate into newly grown layers. Additionally, mismatches in nitrogen concentration between the substrate and growth layers can induce bulk stress due to lattice mismatch \citep{friel2009}. Thus, careful selection and preparation of substrates is critical.

The substrates are CVD single-crystal diamonds with low nitrogen content ([N$_\text{s}^0$]\,$\leq$\,1\,ppm), closely matching the nitrogen concentration of the intended growth layer ($\sim$\,0.8\,ppm). Substrates are selected based on low birefringence, indicative of a low density of dislocations that are likely to thread into the growth. Substrates have typical dimensions of approximately 3.5$\times$3.5$\times$0.3\,mm$^3$, with a mechanically polished $\{100\}$ surface prepared by standard scaife techniques \citep{schuelke2013,hird2004}. After approximately 1.2\,mm of growth, the newly grown CVD diamond layers are laser-sliced to detach from substrates, and only the central 3$\times$3\,mm$^2$ region is retained, avoiding dislocations propagating from substrate edges.

Following processing, electron irradiation and annealing is performed to convert grown-in [N$_\text{s}^0$] defects into NV centers. Samples are irradiated using a 1.5\,MeV electron beam at an estimated dose of $\sim$\,4.8$\times10^{17}$\,cm$^{-2}$, targeting approximately 50\% N$_\text{s}^0$-to-NV conversion rate. Annealing is carried out using a previously established ramp profile with a final temperature of 1200\,$^\circ$C for 2 hours \citep{edmonds2021characterisation}.

\subsection{Wide-Field Strain Mapping}
Strain variations across each NV-diamond sample are mapped using CW-ODMR in a wide-field imaging configuration, as previously described in Ref. \cite{roncaioli2024all}. Briefly, a 532\,nm laser beam ($\sim$\,1\,W) illuminates the diamond at a shallow incidence angle.
Before reaching the diamond,
a 30\,mm focal-length cylindrical lens shapes the laser beam into an elliptical profile, measuring $\sim\,10\,\text{mm}$ by $\sim\,700\,\mu\text{m}$, ensuring uniform illumination across the diamond. NV PL is collected using a 0.1NA, 4$\times$ objective and imaged onto a camera.

A bias magnetic field $\sim\,50\,\text{G}$ is applied to the NV-diamond sample under study using a pair of permanent magnets, with the field oriented such that spin resonances from all four NV orientations in the diamond lattice are resolved. This configuration allows extraction of the NV spin-strain coupling terms ($M_z$) following the method described in \cite{Kehayias2019stress}. Microwaves are delivered to the sample using a loop-gap resonator \citep{eisenach2018broadband}. Pixel-by-pixel fitting of the NV Hamiltonian to the CW-ODMR spectra is performed using GPU-accelerated fitting \citep{przybylski2017gpufit}, generating full-diamond strain maps.

\subsection{Confocal Spin and Charge Characterization}
\label{sec:2_4}
NV spin dephasing times ($T_2^*$), charge-state fractions, PL contrast, and experimental overhead (spin initialization and readout) are measured using a custom-built confocal setup. Measurements are performed as functions of optical NV excitation intensity.

To achieve uniform optical NV excitation, the 532\,nm laser beam is focused to a waist diameter of about 15\,$\mu$m by underfilling a 0.75NA, 20$\times$ objective. A pinhole in the detection path restricts NV PL collection to a smaller ($\sim$\,6\,$\mu$m diameter, $\sim\,10\,\mu\text{m}$ depth) region at the center of the beam spot. The optical intensity at the diamond surface is determined from the average intensity within the restricted collection volume, and is controllably varied from approximately $10^{-3}$ to $10^1$\,mW/$\mu$m$^2$.

Microwave pulses for NV spin-state manipulation are delivered via a coaxial loop near the diamond surface, achieving Rabi frequencies $\sim$\,5\,MHz. A bias magnetic field ($\sim$\,20\,G), aligned along a single NV crystallographic axis, is optimized to minimize spatial inhomogeneity within the probed region, similar to \cite{bauch2018ultralong}.

For NV charge-state determination, PL emission is collected through a multimode optical fiber coupled to a spectrometer. Recorded emission spectra are decomposed into NV$^-$ and NV$^0$ spectral profiles \citep{aude2020microwave}. NV$^-$ charge fractions are then quantified using the intensity ratio between NV$^-$ and NV$^0$ PL \citep{Alsid2019nvcreation}.

\section{Results}

\subsection{Synthesized Diamonds}

\begin{figure}[htbp]
    \centering
    \includegraphics[width=\linewidth]{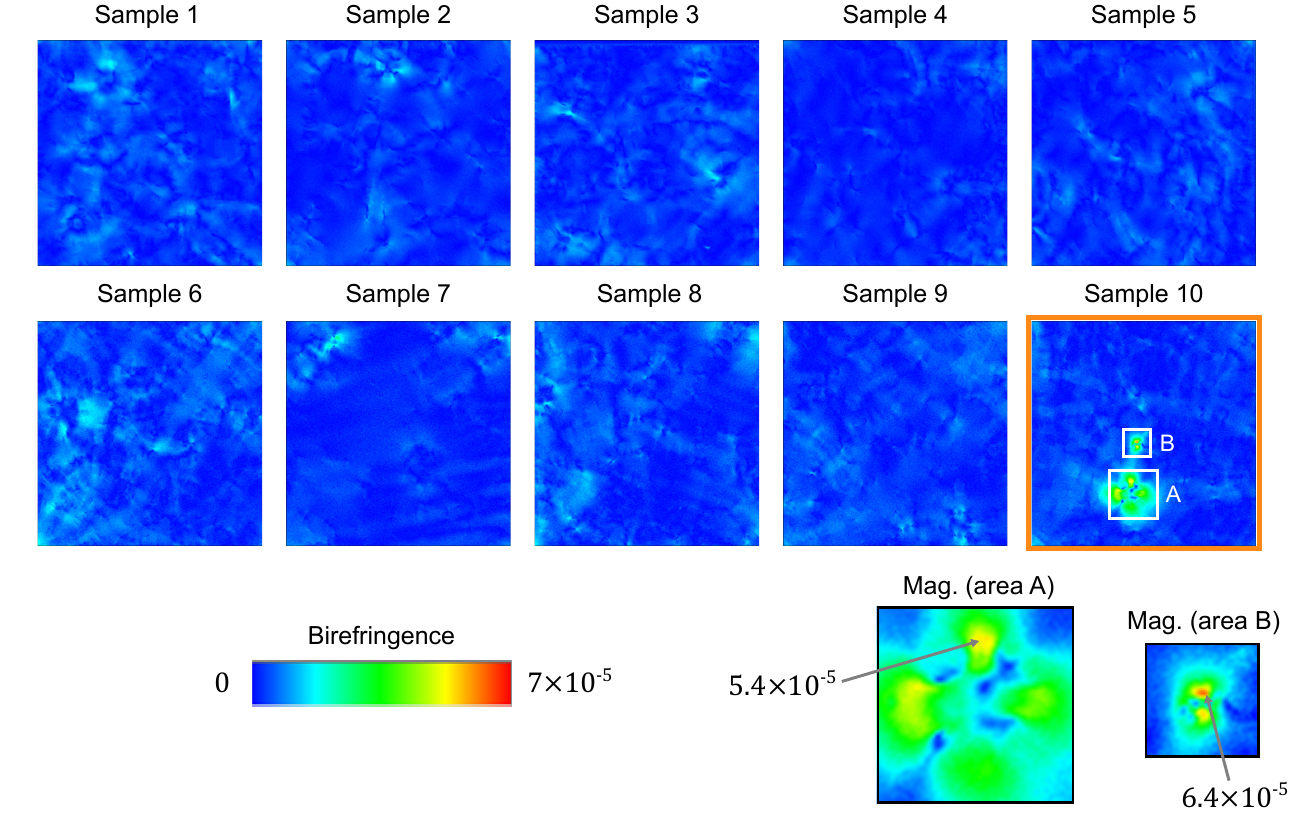}
    \caption{Optical birefringence images of ten low-[N$_\text{s}^0$] diamond samples from a single CVD synthesis run, prior to irradiation and annealing. Sample 10, exhibiting the strongest birefringence features, is selected for subsequent NV-based measurements after irradiation and annealing. Magnified views are shown of two regions of sample 10 with higher levels of birefringence (labeled A and B), with example values of high birefringence indicated.
    } 
    \label{fig:2}
\end{figure}

We first study ten low-[N$_\text{s}^0$] diamonds, grown in a single synthesis run, prior to irradiation and annealing. Measurements of [N$_\text{s}^0$] performed via UV-Vis absorption \citep{edmonds2021characterisation} yield values in the range of $0.73 – 0.88$\,ppm. Optical birefringence imaging at 590\,nm wavelength, performed using techniques outlined in \citep{glazer1996automatic,friel2009}, yields average birefringence ($\Delta n$) across the samples between $2.6\times10^{-6}$ and $5.2\times10^{-6}$ (Fig.~\ref{fig:2}), indicative of generally good strain homogeneity in this diamond cohort \citep{friel2009}. Irradiation and annealing increase the NV concentration in each sample to approximately 0.39(2)\,ppm, as measured by UV-Vis absorption in \citep{edmonds2021characterisation}.

For subsequent characterizations (described below and in the Supplementary Material) we study sample 10, as this sample exhibits higher levels of birefringence from this growth process in certain spatially restricted regions, allowing us to examine both typical behavior for the low-[N$_\text{s}^0$] diamonds (far from the high birefringence regions) as well as the worst-case impact of strain on NV spin dephasing and sensing performance (in the high birefringence regions).

\subsection{NV Spin Dephasing Time and Strain Contribution}

\begin{figure}[htbp]
    \centering
    \includegraphics[width=\linewidth]{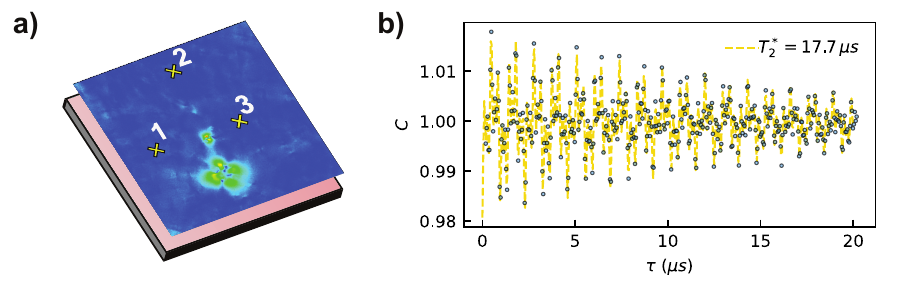}
    \caption{Single-quantum (SQ) Ramsey measurements on diamond sample 10, far from the high birefringence regions. (a) Ramsey measurements performed at three representative locations of low strain, overlaid on the birefringence image from Fig.~\ref{fig:2}. (b) Ensemble NV spin dephasing at location 2. Here the microwave frequency is detuned from the center of the hyperfine-split NV spin transitions, allowing the decay envelope to be separated from hyperfine-induced beating. The extracted ensemble NV dephasing time is $T_2^*=17.7(4)\,\mu\text{s}$.
    }
    \label{fig:3}
\end{figure}

Confocal Ramsey measurements of the single-quantum (SQ) NV ensemble spin dephasing time at three representative low-birefringence locations within sample 10 (Fig.~\ref{fig:3}(a,b)) yield an average value $T_{2,\text{SQ}}^* = 17.5 \pm 1.0\,\mu$s (an additional high-birefringence location measurement is shown in the Supplementary Material). These results are only slightly smaller than the theoretical spin-bath-limited value of approximately 20\,$\mu$s, consistent with NV ensemble spin dephasing in the low-[N$_\text{s}^0$] diamond samples being  primarily limited by intrinsic nitrogen spin-bath interactions (see Supplementary Material). Similar measurements on a previously studied higher-[N$_\text{s}^0$] ($\sim$\,14\,ppm) diamond sample \citep{edmonds2021characterisation} yield $T_{2,SQ}^*\approx1\,\mu\text{s}$. 

\begin{figure}[htbp]
    \centering
    \includegraphics[width=\linewidth]{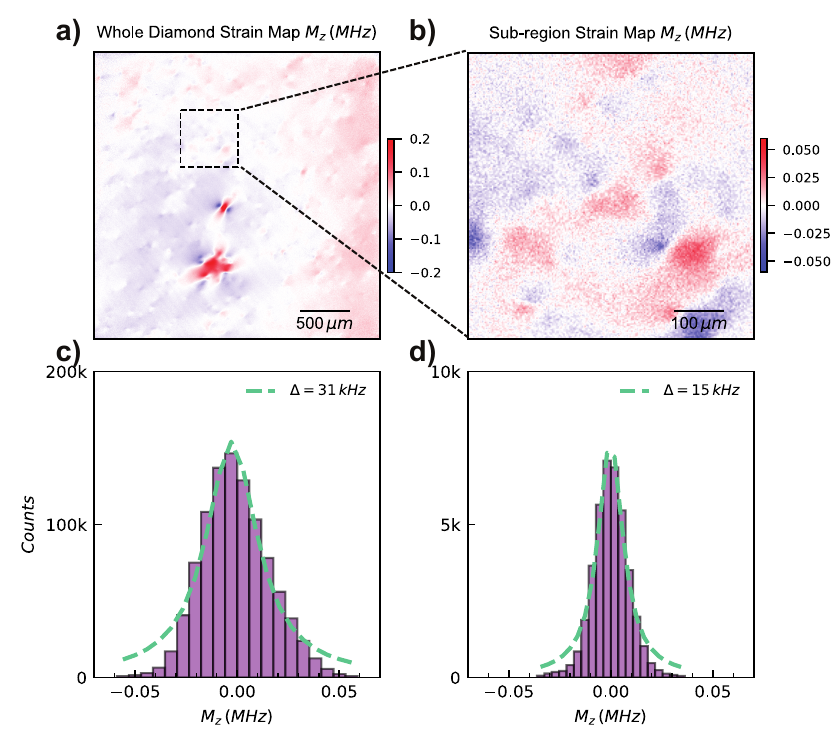}
    \caption{Wide-field imaging of diamond strain for sample 10. (a, b) Spatial maps of strain-induced frequency shifts in NV spin transitions across the entire diamond sample and within a selected sub-area. Strain fields shown are projections onto one NV crystallographic orientation. Each spatial map is mean-subtracted so that the global average is zero, emphasizing relative strain variations relevant to NV spin dephasing. (c, d) Histogram distributions of the strain-induced frequency shifts extracted from the maps in (a) and (b), respectively. Lorentzian fits to the histograms are also shown, from which linewidths (FWHM) $\Delta$ are determined.
    } 
    \label{fig:4}
\end{figure}

Figure \ref{fig:4}(a) shows the strain-induced frequency shifts ($M_{z}$) in NV spin resonances measured across sample 10 for one NV crystallographic orientation (maps for all four NV orientations are available in the Supplementary Material). The distribution of frequency shifts provides a quantitative estimate of the strain-limited NV electronic spin dephasing time \citep{barry2020report}, calculated as $T_{2,\text{strain}}^* = (\pi \Delta)^{-1}$, where $\Delta$ is the full width at half maximum (FWHM) of the distribution. Histogram analysis of strain across the diamond (Fig. \ref{fig:4}(c)) yields $\Delta\approx31\,\text{kHz}$, corresponding to $T_{2,\text{strain}}^* \approx 10\,\mu$s. These wide-field measurements integrate strain variations through the full thickness and across the full diameter of the diamond, overestimating strain effects when considering smaller sensing volumes. For example, histogram analysis of strain in a smaller, low-birefringence region ($\sim\,500\,\mu\text{m}$ diameter) of sample 10 yields a narrower FWHM linewidth of about 15\,kHz and hence $T_2^*\approx20\,\mu\text{s}$ (Fig.~\ref{fig:4}(b,d)), indicating reduced strain inhomogeneity in this smaller region. Extrapolating to the volume probed with confocal measurements, we estimate a further order-of-magnitude reduction in strain inhomogeneity and hence a corresponding modest contribution to the SQ NV ensemble dephasing time $T_{2,SQ}^*$ determined from confocal measurements.

\begin{figure}[htbp]
    \centering
    \includegraphics[width=\linewidth]{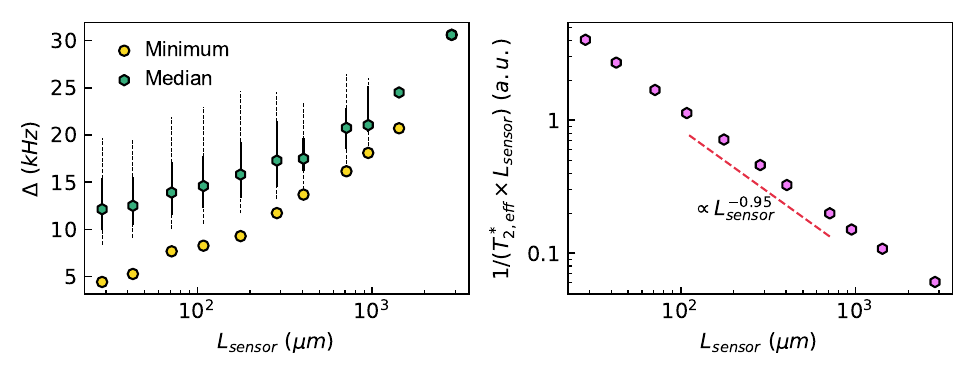}
    \caption{(a) Lorentzian-fitted FWHM ($\Delta$) of strain-induced frequency shifts as a function of sensor size $L_{\text{sensor}}$ for sample 10. Sensor size is systematically varied by partitioning the full-diamond strain map (Fig.~\ref{fig:4}(a)) into smaller square sub-regions. Statistical analysis of $\Delta$ across sub-regions is presented, including minimum and median values. When more than five sub-regions are available, interquartile ranges are included: 25-75\% (vertical solid lines) and 10-90\% (vertical dotted lines). (b) Scaling of effective Ramsey DC magnetometry sensitivity metric with increasing sensor size. In the ideal scenario, where increasing the sensing volume introduces no additional NV ensemble dephasing -- and thus no reduction in Ramsey $T_2^*$ -- the sensitivity metric scales inversely with sensor size. Here, fitting a power-law function to the effective sensitivity metric vs. $L_\text{sensor}$ yields nearly inverse linear scaling behavior, indicating minimal strain limitation on the Ramsey dephasing time for bulk NV ensemble sensing using the present low-[N$_\text{s}^0$] diamonds. The overall low-strain effect of these diamonds applies even to sample 10 with its isolated regions of relatively high strain.} 
    \label{fig:5}
\end{figure}

To evaluate the feasibility of scaling sensor sizes for bulk NV magnetometry, we analyze how strain-induced linewidth broadening changes with sensing region size ($L_\text{sensor}$). We partition the wide-field strain map show in Fig.~\ref{fig:4}(a) into square sub-regions of varying areas ($\sim$\,30\,$\mu$m$^2$ to $\sim$\,3000\,$\mu$m$^2$), computing strain-induced frequency shift distributions for each sub-region, as in Fig.~\ref{fig:4}(c,d). We find that the median of the distribution FWHM ($\Delta$) varies only about a factor of three over this range (Fig.~\ref{fig:5}(a)), indicating modest strain heterogeneity over large lengthscales within the diamond sample.

To illustrate how size-dependent strain-induced dephasing affects NV ensemble sensor performance, we compute an effective Ramsey DC magnetometry sensitivity metric $(T_{2,\text{eff}}^* \times L_\text{sensor})^{-1}$, with $T_{2,\text{eff}}^*$ derived from Eq.~\ref{eq:2} using the median $\Delta$ corresponding to each sensor size. Assuming a constant NV ensemble sensor thickness across all regions of the diamond sample, sensor volume (and hence the number of NVs being probed) scales as $L_\text{sensor}^2$, resulting in inverse linear scaling of the effective sensitivity metric with $L_\text{sensor}$ in the ideal case of no additional NV ensemble dephasing with increasing $L_\text{sensor}$. The observed nearly inverse linear scaling of the effective sensitivity metric for sample 10 (Fig.~\ref{fig:5}(b)) indicates minimal reduction in Ramsey $T_2^*$ on lengthscales $\gtrsim\,30\,\mu\text{m}$, suggesting sensor-size scaling remains a viable strategy for improved bulk NV magnetometry using this low-[N$_\text{s}^0$] diamond material.

\subsection{Photoluminescence Contrast, Overhead Time, and Charge Fraction}

\begin{figure}[htbp]
    \centering
    \includegraphics[width=6.5in]{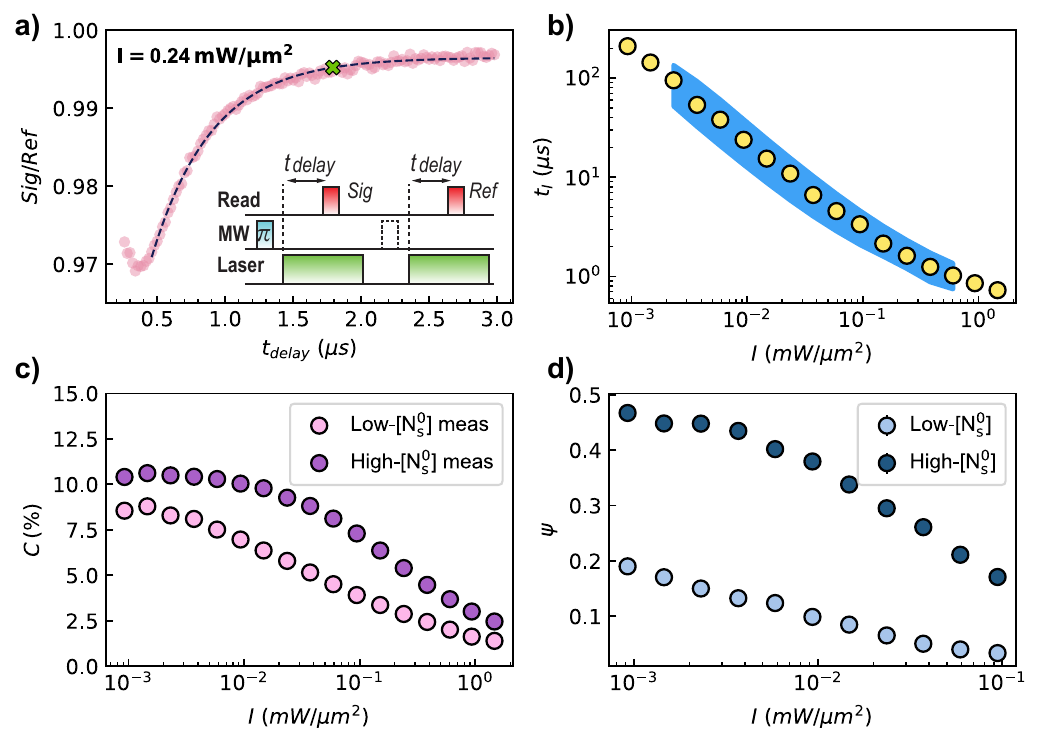}
    \caption{NV spin polarization dynamics, initialization time, PL contrast, and charge fraction as functions of laser excitation intensity. (a) Measured NV spin polarization dynamics and schematic measurement protocol (inset). A 1-ms laser pulse (not shown) initializes the NV ensemble into the $m_s=0$ state prior to a microwave (MW) $\pi$ pulse that initializes the NV ensemble into $m_s = -1$.  A laser pulse then begins to repolarize the NV spins, with PL readout (Sig) acquired after a delay time $t_\text{delay}$. The measurement protocol is then repeated without the MW $\pi$ pulse, and a reference PL measurement (Ref) is acquired. Example PL contrast measurement (Sig/Ref) at a laser intensity of 0.24\,mW/$\mu\text{m}^2$, shown as a function of $t_\text{delay}$. Spin initialization time $t_I$ (green cross) is extracted by fitting an exponential decay to the PL contrast data following its peak; we set $t_I$ as the delay at which contrast decays to $1/e^3$ of its peak value, corresponding to repolarization of $\sim$\,95\% of the ensemble NV spins. With this measurement protocol, the readout time is included in the reported initialization time.
    (b) Measured NV initialization time ($t_I$) as a function of laser intensity. The shaded blue region represents the range of initialization times predicted based on a five-level model of NV photophysics, with saturation intensities of 1\,mW/$\mu\text{m}^2$ and 3\,mW/$\mu\text{m}^2$ used to produce the lower and upper bounds, respectively (see Supplementary Material). (c) Measured peak PL contrast for low- and high-[N$_\text{s}^0$] diamonds as a function of laser intensity.
    (d) Experimentally-determined NV charge-state fraction $\psi$, defined as [NV$^{-}$]/([NV$^{-}$]+[NV$^{0}$]), as a function of laser intensity for both low- and high-[N$_\text{s}^0$] diamond samples. NV PL spectra used for decomposition analysis (see the main text) are collected using a 550\,nm long-pass filter.
    } 
    \label{fig:6}
\end{figure}

We characterize NV photoluminescence (PL) spin-state contrast ($C$) and spin initialization time ($t_I$) as functions of optical excitation intensity. Figure~\ref{fig:6}(a) illustrates the protocol used to measure PL contrast and NV spin polarization dynamics. After initializing NV spins into either $m_s = 0$ or $m_s = -1$, we vary the delay ($t_\text{delay}$) to NV spin-state  readout. PL contrast initially reaches a maximum before decreasing as NV spins repolarize to $m_s = 0$, allowing simultaneous extraction of $C$ and $t_I$  from a single measurement (Fig.~\ref{fig:6}(b,c)). To validate the experimental calibration of laser power to incident optical excitation intensity for the measured NV ensemble, we simulate NV spin polarization dynamics using a five-level photophysics model \citep{schloss2019optimizing} (see Supplementary Material). Simulated initialization times ($t_I$) closely match experimental data across a wide range of optical intensities (Fig.~\ref{fig:6}(b)), confirming accurate calibration of illumination conditions within the probed region.

Observed PL contrast degradation at higher optical intensities (Fig.~\ref{fig:6}(c)) is primarily attributed to increased NV$^-$ ionization \citep{aslam2013photo, manson2005photo}, raising the NV$^0$ fraction and associated background PL. Similar contrast reduction at high optical intensity is observed in the higher-[N$_\text{s}^0$] ($\sim$\,14\,ppm) diamond sample, but is partially mitigated in this sample due to increased nitrogen donor concentration stabilizing NV$^-$. Direct measurements of the NV charge-state fraction ($\psi$) using a PL emission spectrum decomposition technique \citep{aude2020microwave, Alsid2019nvcreation} reveal decreasing NV$^-$ fraction with increasing optical intensity (Fig.~\ref{fig:6}(d)). The charge-state determination is limited to intensities $\lesssim$\,10$^{-1}$\,mW/$\mu$m$^2$, where the relative PL emission ratio between NV$^0$ and NV$^{-}$ is reliably determined below NV saturation conditions \citep{Alsid2019nvcreation}. Measurement of $\psi$ also  shows substantially lower NV$^-$ fraction in low-[N$_\text{s}^0$] compared to high-[N$_\text{s}^0$] diamond (Fig.~\ref{fig:6}(d)). However, the NV$^0$ PL emission intensity is approximately 2.5$\times$ dimmer than that for NV$^-$ under typical optical excitation \citep{Alsid2019nvcreation}, and spectral filtering techniques used in NV experiments (passing PL $>$\,647\,nm) further reduce the effect of NV$^0$ PL on observed NV-spin-state contrast. 

\section{Discussion}
The availability of low-[N$_\text{s}^0$], NV-enriched, and $^{12}$C isotopically purified diamond material with good strain homogeneity presents opportunities for NV ensemble sensing applications. In particular, our systematic characterization of NV spin and charge-state properties for both low- and high-[N$_\text{s}^0$] samples enables a comparison of their expected photon-shot-noise-limited DC magnetic sensitivity as a function of optical excitation intensity.

\begin{figure}[htbp]
    \centering
    \includegraphics[width=\linewidth]{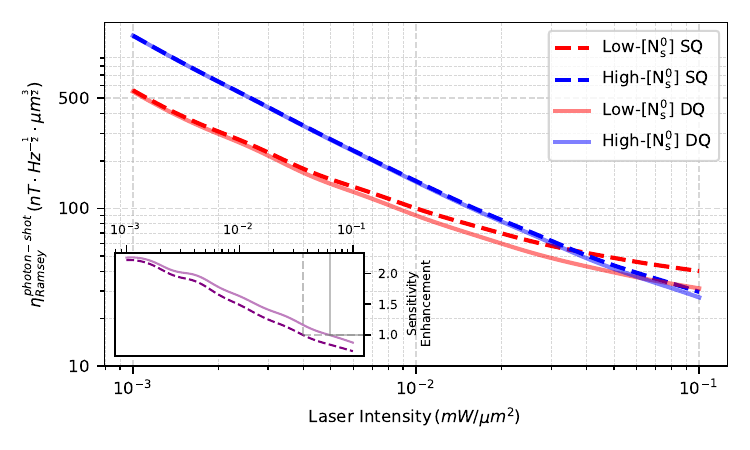}
    \caption{Estimated volume-normalized photon shot-noise limited DC magnetic sensitivity for low- and high-[N$_\text{s}^0$] diamond samples as functions of optical excitation intensity. We assume a photon emission rate of 30\,kcps per single NV at 1\,mW/$\mu$m$^2$ illumination intensity. Sensitivity-relevant parameters (see Eq.~\ref{eq:1}), including the Ramsey dephasing times ($T_2^*$ for both single- and double-quantum measurements), spin-state-dependent PL contrast ($C$), and experimental overhead time ($t_O$), are directly measured as functions of laser intensity. Nitrogen concentration and NV charge-state fraction, experimentally determined using the UV-Vis absorption method (Section \ref{sec:2_2}), are used to estimate the number of NV$^-$ centers contributing to sensing at a given laser intensity. 
    The inset shows the ratio of estimated sensitivity between high-[N$_\text{s}^0$] and low-[N$_\text{s}^0$] samples as a function of optical excitation intensity, for SQ (solid) and DQ (dashed) Ramsey protocols, highlighting the relative advantage of low-nitrogen material at moderate to low optical excitation intensities.
    } 
    \label{fig:7}
\end{figure}

Figure~\ref{fig:7} summarizes volume-normalized DC sensitivity estimates for both diamond types, effectively replacing the NV$^-$ number ($N$ in Eq.~\ref{eq:1}) with the NV$^-$ concentration [NV$^-$]. At moderate to low optical excitation intensities (below $\sim$\,0.03\,mW/$\mu$m$^2$), the low-[N$_\text{s}^0$] sample exhibits better (i.e., smaller) estimated sensitivity compared to the high-[N$_\text{s}^0$] sample, despite its lower NV density and reduced NV$^-$ charge fraction. This sensitivity advantage is primarily due to the low-[N$_\text{s}^0$] sample's longer NV spin dephasing time $T_2^*$ (see Eq.~\ref{eq:1}), which improves the sensing duty cycle, particularly when experimental overhead times associated with initialization and readout are substantial. At higher optical excitation intensities, reduced overhead time narrows the difference in sensing duty cycles between the low- and high-[N$_\text{s}^0$] diamonds, making differences in the NV charge-state fraction and effective NV$^-$ density the dominant factors for sensitivity.

Double-quantum (DQ) Ramsey protocols applied to NV ensembles typically provide DC magnetic field sensitivity enhancement via a doubled effective gyromagnetic ratio ($\Delta m_s=2$ in Eq.~\ref{eq:1}) and robustness against strain-induced NV spin dephasing \citep{Hart2021-4Ramsey}. However, in diamond samples with good strain homogeneity -- such as the low-[N$_\text{s}^0$] diamond studied here -- or for small interrogation volumes with minimal strain variation, the advantages of DQ sensing strongly depend on the relationship between the spin dephasing time ($T_2^*$) and experimental overhead time ($t_O$). Due to increased sensitivity to spin-bath noise, DQ dephasing times are typically about half those of single-quantum (SQ) protocols. Thus, significant DQ sensitivity improvements primarily arise when dephasing times approach or exceed the overhead time ($T_2^*\gtrsim t_O$), as sensitivity scales linearly with $\Delta m_s$ but sublinearly with $T_2^*$. This criterion can be readily satisfied by low-[N$_\text{s}^0$] diamonds, whereas the shorter dephasing times ($\lesssim$\,1\,$\mu$s) characteristic of higher-[N$_\text{s}^0$] diamonds typically result in linear scaling of sensitivity with $T_2^*$.

We note that diamond sample 10 examined in this study (Fig.~\ref{fig:2}) exhibited the highest level of birefringence among the ten low-[N$_\text{s}^0$] plates initially characterized, with increased birefringence concentrated in two regions. The pattern is indicative of dislocation bundles, most likely originating from residual damage due to mechanical polishing of the CVD substrates, rather than from the underlying bulk dislocation content of the substrates. If further strain reduction is required, alternative polishing techniques and enhanced substrate characterization for subsurface damage can be implemented \citep{graziosi2023}.

Our results provide practical guidance for selecting NV ensemble diamond samples for use as DC magnetic field sensors under realistic optical excitation constraints. Applications with limited budgets for size, weight, power, and cost (SWaP-C), or biological sensing that aims to minimize phototoxicity caused by intense green laser illumination \citep{laissue2017assessing}, will benefit from low-[N$_\text{s}^0$] diamonds. However, limitations of low-nitrogen diamonds under high-intensity illumination highlight the continued need to optimize NV$^-$ charge-state stability in future diamond material engineering efforts.

\section{Conclusion}
We synthesized and systematically characterized low-[N$_\text{s}^0$] ($\sim$\,0.8\,ppm), $^{12}$C-purified diamond samples tailored for pulsed DC magnetometry applications using NV ensembles. Our results demonstrate that carefully engineered low-[N$_\text{s}^0$] diamond grown on CVD substrates can achieve high spatial strain homogeneity and spin-bath-limited NV spin dephasing times. Controlled irradiation and annealing enriches NV concentration while retaining relatively stable NV$^-$ charge-state fraction under optical illumination. This combination of properties yields enhanced photon-shot-noise-limited DC magnetic sensitivity, surpassing higher-[N$_\text{s}^0$] diamond sensors at moderate to low optical excitation intensities.

These findings provide practical benchmarks and guidance for selecting NV-diamond sensors based on available optical power and application-specific requirements. Specifically, low-[N$_\text{s}^0$] diamonds are advantageous for applications requiring low optical illumination intensity, e.g., providing reduced size, weight, power, and cost (SWaP-C) and minimal phototoxicity. Additionally, the NV spin dephasing times achievable in low-[N$_\text{s}^0$] diamonds should benefit from spin-bath driving techniques \citep{barry2024sensitive,bauch2018ultralong}, potentially enabling DC magnetometry with advanced readout protocols \citep{arunkumar2023quantum}, further extending the sensitivity and application scope of NV-based quantum sensors.

\section*{Conflict of Interest Statement}
R.L.W. is a founder of and advisor to companies that are developing and commercializing NV sensing technology. These relationships are disclosed to and managed by the University of Maryland Conflict of Interest Office. These companies did not have any relationship with this study.

\section*{Author Contributions}
JT: Conceptualization, Formal Analysis, Investigation, Methodology, Software, Validation, Visualization, Writing -- original draft, Writing -- review \& editing; CAR:  Formal Analysis, Investigation, Methodology, Resources, Software, Validation, Writing -- original draft, Writing -- review \& editing; AME: Formal Analysis, Investigation, Methodology, Resources, Software, Validation, Visualization, Writing -- original draft, Writing -- review \& editing; AD: Investigation, Writing -- review \& editing; CAH: Conceptualization, Resources, Validation, Writing -- review \& editing; MLM: Conceptualization, Resources, Writing -- review \& editing; RLW: Conceptualization, Funding acquisition, Project administration, Resources, Supervision, Writing - review \& editing.

\section*{Funding}
This work is supported by, or in part by, the U.S. Army Research Laboratory, under Contract No. W911NF2420143; the U.S. Army Research Office, under Grant No. W911NF2120110; and the University of Maryland Quantum Technology Center.

Research was sponsored by the Army Research Office and was accomplished under W911NF2420143 Cooperative Agreement Number W911NF-24-2-0143. The views and conclusions contained in this document are those of the authors and should not be interpreted as representing the official policies, either expressed or implied, of the Army Research Office or the U.S. Government. The U.S. Government is authorized to reproduce and distribute reprints for Government purposes notwithstanding any copyright notation herein.

\section*{Supplemental Data}
See the Supplementary Material.

\section*{Data Availability Statement}
The data that support the findings of this study are available upon reasonable request from the authors.

\bibliographystyle{Frontiers-Harvard}
\bibliography{reference}

\clearpage               
\beginsupplement         

\section*{Supplementary Material}
\addcontentsline{toc}{section}{Supplementary Material} 
\markboth{Supplementary Material}{Supplementary Material} 

\section{Strain and Stress Maps}

\begin{figure}[htbp]
\begin{center}
\includegraphics[width=\linewidth]{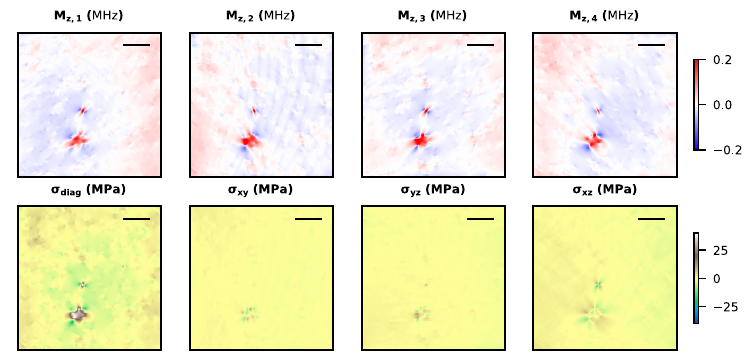}
\end{center}
\caption{Spatial maps of experimentally determined NV spin-strain coupling terms ($M_z$) for each of four NV orientations and derived diamond stress tensors components $\sigma$ across the entire diamond sample 10. Scale bars: $500\,\mu\text{m}$}\label{fig:s1}
\end{figure}

The NV electronic spin Hamiltonian contains spin-strain coupling terms, including longitudinal ($M_z$) and transverse ($M_x, M_y$) components \citep{barfuss2019spin}. Under a strong bias magnetic field ($|\vec{B}|\gtrsim10\,G$), the longitudinal component $M_z$ dominates NV spin resonance frequency shifts, and transverse contributions can be neglected \citep{Kehayias2019stress}. By resolving NV spin transitions corresponding to the four NV orientations in the diamond lattice, the spin stress terms $M_{z,i}$ for each orientation $i$ are experimentally determined using continuous-wave optically detected magnetic resonance (CW-ODMR) measurements, as described in \cite{Kehayias2019stress}. Representative spatial maps of these terms across diamond sample 10 are shown in Fig.~\ref{fig:s1}.

Additionally, from these strain measurements, the stress tensor components of the diamond -- comprising diagonal terms ($\sigma_{diag}=\sigma_{xx}+\sigma_{yy}+\sigma_{zz}$) and off-diagonal shear terms ($\sigma_{xy}, \sigma_{yz}, \sigma_{xz}$) -- can be uniquely determined. The calculation method is detailed in Ref. \cite{Kehayias2019stress} and the results are also shown in Fig.~\ref{fig:s1}.

\section{Ramsey Measurement at High-Birefringence Location}
As shown in the main text (Fig. 2), sample 10 contains two regions with elevated birefringence, indicating higher strain variation. To illustrate the effect of such inhomogeneity on NV ensemble properties, we perform a single-quantum (SQ) Ramsey measurement at a spot within area B. The extracted dephasing time is $T_{2,SQ}^* = 5.1(6),\mu\text{s}$, significantly shorter than the spin-bath-limited value. In contrast, a double-quantum (DQ) Ramsey measurement at this same spot, which is insensitive to longitudinal strain and temperature variations \citep{Hart2021-4Ramsey}, yields $T_{2,DQ}^* = 8.6(5),\mu\text{s}$. This value is approximately half of the SQ dephasing time observed in low-birefringence regions (main text, Fig. 3), consistent with the doubled gyromagnetic ratio and increased sensitivity to spin-bath noise of DQ measurements.

\begin{figure}[htbp]
\begin{center}
\includegraphics[width=5in]{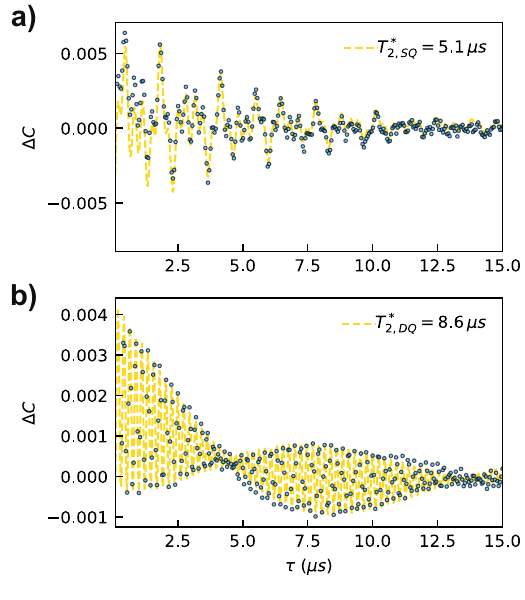}
\end{center}
\caption{Single-quantum (SQ) and double-quantum (DQ) Ramsey measurements of NV ensemble spin dephasing time at a high-birefringence region (area B of sample 10, see main text Fig. 2). The reduced $T_{2,\text{SQ}}^*$ reflects the strong local strain variation, while the DQ protocol mitigates these effects. $\Delta C$ is the change in NV PL contrast.}\label{fig:s2}
\end{figure}

\section{Spin-bath-limited NV Spin Dephasing Time}

The spin-bath-limited NV spin dephasing time ($T_{2,\text{bath}}^*$) can be estimated using the following relation \citep{barry2020report}:

\begin{equation}
\begin{split}
    \frac{1}{T_{2,\text{bath}}^*} &=  \frac{1}{T_{2,\text{[N}_\text{s}^0\text{]}}^*}+\frac{1}{T_{2,\text{[}^{13}\text{C]}}^*}+\frac{1}{T_{2,\text{NV-NV}}^*} \\
    & \approx A_{\text{N}_\text{s}^0}\text{[N}_\text{s}^0\text{]}
    + A_{^{13}\text{C}}\text{[}^{13}\text{C}\text{]}+\zeta_{\parallel}A_{\mathrm{NV}_{||}^-}[\mathrm{NV}_{||}^-]+\zeta_{\nparallel}A_{\mathrm{NV}_{\nparallel}^-}[\mathrm{NV}_{\nparallel}^-],
\end{split}
\end{equation}
where the scaling constants are $ A_{\text{N}_\text{s}^0}=0.101\,\mu\text{s}^{-1}\text{ppm}^{-1}$, $A_{^{13}\text{C}}=0.1\,\text{ms}^{-1}\text{ppm}^{-1}$, $A_{\mathrm{NV}_{||}^-}=0.247\,\mu\text{s}^{-1}\text{ppm}^{-1}$ and $A_{\mathrm{NV}_{\nparallel}^{-}}=0.165\,\mu\text{s}^{-1}\text{ppm}^{-1}$. Here, $[\mathrm{NV}_{||}^-]$ represents the concentration of negatively-charged NV centers (NV$^-$) aligned along the sensing axis, while $[\mathrm{NV}_{\nparallel}^-]$ corresponds to the concentration of NV centers oriented along other crystal axes. For our experimental study, the bias magnetic field aligns with a single NV crystal axis, resulting in $[\mathrm{NV}_{||}^-]\approx\frac{1}{3}[\mathrm{NV}_{\nparallel}^-]$. For the sensing axis and other axes, the dimensionless correction factor $\zeta$ accounts for the incomplete polarization of NV$^-$ spins into the $m_s=0$ state. We optimize laser polarization such that NV centers aligned along two axes are predominantly excited, with one orientation used for sensing. Thus, we approximate $\zeta_{[\mathrm{NV}_{\parallel}^-]}\approx0$. Determining $\zeta_{[\mathrm{NV}_{\nparallel}^-]}$ is complicated due to unequal laser excitation intensities and polarizations along the three non-sensing NV orientations under epi-illumination conditions; as a conservative estimate for the NV-NV interaction-limited $T_{2,\text{NV-NV}}^*$, we assume partial non-sensing NV spin polarization with $\zeta_{\nparallel}=0.5$.

Before electron irradiation and annealing of the ten low-[N$_\text{s}^0$] diamond samples, the substitutional nitrogen concentration is measured to be $\text{[N}_\text{s}^{0, \text{as-grown}}\text{]}\approx0.8\,\text{ppm}$ via UV-Vis absorption \citep{edmonds2021characterisation}. After processing, the total NV concentration is 0.39(2)\,ppm determined by UV-Vis absorption \citep{edmonds2021characterisation}, with approximately 20\% of NVs in the negatively charged (NV$^-$) state (as determined from  the PL emission spectrum decomposition technique, see main text). Formation of NV centers consumes substitutional nitrogen, with typically one nitrogen atom per NV$^0$ and two per NV$^-$ center \citep{barry2020report}. Therefore, the post-treatment substitutional nitrogen concentration $\text{[N}_\text{s}^0\text{]}\approx0.35\,\text{ppm}$. Residual $^{13}$C concentration is 0.01\% (equivalent to 108\,ppm) given the isotopic purification used in the CVD process. Using these values, we estimate the spin-bath-limited NV spin dephasing time to be $T_{2,\text{bath}}^*\approx 20\,\mu\text{s}$, slightly larger than the experimentally measured value for a single quantum (SQ) Ramsey measurement (see main text).

\section{Increasing Sensor Size for Bulk NV Ensemble Sensing}
For photon-shot-noise-limited Ramsey-based DC magnetic sensing for an NV ensemble (main text Eq.~1), optimal sensitivity typically occurs when the NV spin free-precession time ($\tau$) is close to the spin dephasing time $T_2^*$ (i.e., $\tau\approx T_2^*$) \citep{barry2020report}. Consequently, the relationship between sensitivity ($\eta$) and spin dephasing time can be approximated as:
\begin{equation}
    \eta\propto{\frac{\sqrt{T_2^*+t_O}}{T_2^*}},
\end{equation}
In the limit of large experimental overhead time ($t_O\gg T_2^*$), sensitivity scales inversely with dephasing time ($\eta\propto 1/T_2^*$). Conversely, when overhead time is negligible ($t_O\ll T_2^*$), sensitivity scales with the inverse square root of dephasing time ($\eta\propto 1/\sqrt{T_2^*}$)

Thus, when evaluating the impact of spatially varying diamond strain on sensitivity through its effect on spin dephasing time (main text Eq. 2), the most conservative scenario is considered. In this scenario, any degradation in dephasing time due to higher strain variation across an increased sensor volume directly worsens sensitivity, following the inverse-linear scaling relationship ($\eta\propto 1/T_2^*$).

\section{NV spin polarization dynamics modeling}

We model the NV spin population dynamics under green laser excitation using a simplified five-level system \citep{schloss2019optimizing}. This modeling supports our experimental calibration and and conversion of laser power to incident optical excitation intensity by simulating NV spin initialization times as functions of laser intensity.

\begin{figure}[htbp]
\begin{center}
\includegraphics[width=5in]{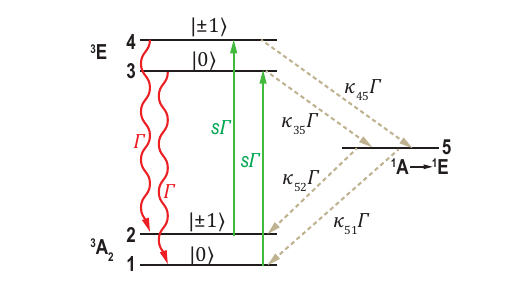}
\end{center}
\caption{Schematic of the simplified five-level NV photophysics model used in spin polarization dynamics simulations. }\label{fig:s3}
\end{figure}

The NV electronic and spin energy levels included in the simulation are illustrated schematically in Fig.~\ref{fig:s3}. The spin-conserving radiative decay rate is set to $\Gamma=0.67\,\mu\text{s}^{-1}$ \citep{schloss2019optimizing}. Transitions between excited and singlet states, as well as subsequent transitions to the ground state, are characterized by decay rates given by $\kappa_{ij}\Gamma$, where $i,j$ denote the initial and final states, respectively. The optical excitation rate from the ground to the excited state is modeled as $s\Gamma$, with saturation parameter $s=I/I_{sat}$, where $I$ is the laser intensity and $I_{sat}$ is the NV saturation intensity. The saturation intensity is taken to be approximately $1-3\,\text{mW}/\mu\text{m}^2$ \citep{barry2020report}, providing bounds for our simulations. The rate-constant prefactors $\kappa$ used in the model are averaged from experimental values summarized in  Ref.~\citep{schloss2019optimizing}: $\kappa_{45}=1, \kappa_{35}=1/7, \kappa_{52}=1/50, \kappa_{51}=1/25$.

For the experimental measurement protocol described in the main text (Fig. 6), we simulate NV spin population dynamics starting from the initial states $m_s=0$ and $m_s=\pm1$. The following system of coupled first-order differential equations describes the population $n_i$ of each of the five modeled NV states: 
\begin{equation}
    \begin{split}
        &dn_1/dt=\Gamma(-sn_1(t)+n_3(t)+\kappa_{51}n_5(t))\\
        &dn_2/dt=\Gamma(-sn_2(t)+n_4(t)+\kappa_{52}n_5(t))\\
        &dn_3/dt=\Gamma(sn_1(t)-(1+\kappa_{35})n_3(t))\\
        &dn_4/dt=\Gamma(sn_2(t)-(1+\kappa_{45})n_4(t))\\
        &dn_5/dt=\Gamma(\kappa_{35}n_3(t)+\kappa_{45}n_4(t)-(\kappa_{51}+\kappa_{52})n_5(t)).
    \end{split}
\end{equation}
The simulated PL emission rate is calculated as: 
\begin{equation}
    R(t)=\Gamma(n_3(t)+n_4(t)),
\end{equation}
and is further processed by a low-pass filter (4th-order Butterworth with -3dB cutoff at 1.7\,MHz) to mimic the effect of data acquisition hardware used experimentally (NI DAQ USB-6363).

The stimulated PL contrast is determined by normalizing the PL trace for spins initialized in the $m_s=\pm1$ state by the PL for spins initialized in the $m_s=0$ state. An exponential fit is applied to the PL contrast decay, with the initialization time ($t_I$) defined as the delay time ($t_{delay}$) at which the contrast decays to $1/e^3$ of its peak value, consistent with the experimental approach described in the main text.



\end{document}